\documentclass[%
reprint,
superscriptaddress,
amsmath,
amssymb,
nature,
]{revtex4-2}
\usepackage[colorlinks = true,
            linkcolor = blue,
            urlcolor  = red,
            citecolor = blue,
            anchorcolor = blue]{hyperref}

\usepackage{graphicx}
\usepackage{dcolumn}
\usepackage{bm}
\usepackage{lineno}
\usepackage{url}
\usepackage{ulem}

\newcommand{\orcid}[1]{\href{https://orcid.org/#1}{\textcolor[HTML]{A6CE39}{\aiOrcid}}}

\usepackage{glossaries}
\newacronym{SSH}{SSH}{Su-Schrieffer–Heeger}
\newacronym{HMC}{HMC}{hybrid Monte Carlo}
\newacronym{CDW}{CDW}{charge-density-wave}
\newacronym{DFPT}{DFPT}{density functional perturbation theory}
\newacronym{DFT}{DFT}{density functional theory}

\makeatletter
\newsavebox{\@brx}
\newcommand{\llangle}[1][]{\savebox{\@brx}{\(\m@th{#1\langle}\)}%
  \mathopen{\copy\@brx\kern-0.5\wd\@brx\usebox{\@brx}}}
\newcommand{\rrangle}[1][]{\savebox{\@brx}{\(\m@th{#1\rangle}\)}%
  \mathclose{\copy\@brx\kern-0.5\wd\@brx\usebox{\@brx}}}
\makeatother

\begin{document}

\preprint{}

\title{A hybrid {Monte Carlo} study of 
bond-stretching electron-phonon interactions and charge order in BaBiO$_3$}
\author{Benjamin Cohen-Stead}
\affiliation{Department of Physics and Astronomy, The University of Tennessee, Knoxville, Tennessee 37996, USA}
\affiliation{Institute for Advanced Materials and Manufacturing, The University of Tennessee, Knoxville, Tennessee 37996, USA\looseness=-1}

\author{Kipton Barros}
\affiliation{Theoretical Division and CNLS, Los Alamos National Laboratory, Los
Alamos, New Mexico 87545, USA}

\author{Richard Scalettar}
\affiliation{Department of Physics and Astronomy, University of California, Davis, California 95616, USA}

\author{Steven Johnston}
\email{sjohn145@utk.edu}
\affiliation{Department of Physics and Astronomy, The University of Tennessee, Knoxville, Tennessee 37996, USA}
\affiliation{Institute for Advanced Materials and Manufacturing, The University of Tennessee, Knoxville, Tennessee 37996, USA\looseness=-1}

\date{\today}

\begin{abstract}
The relationship between electron-phonon ($e$-ph) interactions and charge-density-wave (CDW) order in the bismuthate family of high-temperature superconductors remains unresolved. We address this question using nonperturbative hybrid Monte Carlo calculations for the parent compound BaBiO$_3$. Our model includes the Bi $6s$ and O $2p_\sigma$ orbitals and coupling to the Bi-O bond-stretching branch of optical phonons via modulations of the Bi-O hopping integral. We simulate three-dimensional clusters of up to 4000 orbitals, with input model parameters taken from  {\it ab initio} electronic structure calculations and a phonon energy $\hbar\Omega_0 = 60$~meV. Our results demonstrate that the coupling to the bond-stretching modes is sufficient to reproduce the CDW transition in this system, despite a relatively small dimensionless coupling. We also find that the transition deviates from the weak-coupling Peierls' picture. This work demonstrates that off-diagonal $e$-ph interactions in orbital space are vital in establishing the bismuthate phase diagram.
\end{abstract}

\maketitle
\section{Introduction}

The barium bismuthate families of superconductors Ba$_{1-x}$K$_x$BiO$_3$ and BaBi$_{1-x}$Pb$_x$O$_3$ exhibit high-temperature (high-$T_\mathrm{c}$) superconductivity in proximity to a \gls*{CDW} insulating phase \cite{MattheissPRB1988, CavaNature1988, Sleight2015}. The superconducting transition temperature in these materials is largest in the Ba$_{1-x}$K$_x$BiO$_3$ family with $T_\mathrm{c} \approx 32$~K for $x \approx 0.35$, which is a record among non-cuprate oxides. These materials remain of fundamental interest due to possible analogies with spin-density-wave-driven exotic superconductivity in unconventional superconductors; nevertheless, 
the underlying pairing mechanism and its relationship to the proximate \gls*{CDW} phase remains unsettled. Resolving this question is pertinent with the discovery of superconductivity in related (Ba,K)SbO$_{3}$ \cite{kim2021discovery} and theoretical proposals for  superconductivity in other $s$-$p$ $ABX_3$ perovskites 
~\cite{BenamPRB2021}. 

The bismuthates have no magnetic order in their phase diagram \cite{Sleight2015}, which speaks against unconventional spin-fluctuation-based pairing mechanisms proposed for high-$T_\mathrm{c}$ superconductors~\cite{ScalapinoRMP}. 
Moreover, the extended nature of the Bi $6s$ orbitals and their significant overlap with the neighboring oxygen atoms suggests that strong electronic correlations play a minor role in these materials ~\cite{MattheissPRB1983, HarrisonPRB2006, PlumbPRL2016, FoyevtsovaPRB2015}. 
These aspects make the bismuthates a compelling platform for studying high-$T_\mathrm{c}$ superconductivity in proximity to a charge-ordered state, using uncorrelated models free from issues like the fermion sign problem. Indeed, there has been a recent resurgence of interest in these compounds, using a host of modern thin film syntheses \cite{PlumbPRL2016, Jiao2018, WenPRL2018}, spectroscopic \cite{PlumbPRL2016, NicolettiPNAS2017, WenPRL2018}, and theoretical \cite{FranchiniPRL2009, NourafkanPRL2012, FoyevtsovaPRB2015, NaamnehPreprint2018, LiPRL2019, LiQM2020, JiangPRB2021} techniques. 

In analogy to the doped Mott insulator picture for the cuprates, several leading hypotheses for the pairing mechanism draw insight from the nature of the neighboring \gls*{CDW} phase. 
Bi$^{4+}$ is a skipped oxidation state that does not typically appear in nature. 
Early interpretations of the insulating phase, therefore, proposed that the would-be Bi$^{4+}$ ions undergo charge disproportionation to  (Bi$^{4+\delta}$Bi$^{4-\delta}$) \cite{Cox1976, RicePRL1981, VarmaPRL1988}. 
In the low-temperature mixed valant phase, Bi-O bond distances then relax in response to Bi charge, resulting in a $\mathbf{q}=(\pi,\pi,\pi)a^{-1}$ breathing distortion of the oxygen atoms and an insulating state. In this case, it has been proposed that the displacement of the highly polarizable O atoms towards the Bi atoms generates a $-U$ center on the Bi site~\cite{HarrisonPRB2006}. This scenario would then naturally explain the superconducting state upon doping~\cite{RicePRL1981, VarmaPRL1988, ScalettarPRL1989, FontenelePRB2022}. 
Spectroscopic measurements, however, do not support this charge disproportionation view~\cite{Orchard1977, WertheimPRB1982, PlumbPRL2016}. This picture is also at odds with the significant hybridization between the Bi $6s$ and O $2p$ orbitals and {\it ab initio} electronic structure calculations \cite{FranchiniPRL2009, FoyevtsovaPRB2015, PlumbPRL2016}, which find significant oxygen character for the states near the Fermi level in the metallic phase. As an alternative, it has been argued that the bismuthates are best viewed as negative charge transfer materials~\cite{ZSA,Khomskii}, where the system ``self-dopes'' and transfers one of the Bi holes to a ligand oxygen molecular orbital with  $A_\mathrm{1g}$ symmetry  \cite{IGNATOV2000332, FoyevtsovaPRB2015, KhazraiePRB2018}. 
Recent ARPES data for the parent compound support this view of the electronic structure \cite{PlumbPRL2016}. In this scenario, the $A_\mathrm{1g}$ ligand holes couple strongly to the breathing motion of the lattice \cite{KhazraiePRB2018}, which induces the insulating phase. Upon doping, the $e$-ph coupling and potential polaronic effects generate the subsequent pairing \cite{KhazraiePRB2018, LiQM2020, JiangPRB2021}. 

Finally, it has been argued that the long-range Coulomb interaction in these materials can enhance the $e$-ph interaction, leading to a more ``conventional'' BCS-like mechanism \cite{YinPRX2013, WenPRL2018}. The standard BCS scenario was originally ruled out because a na{\"i}ve estimate for $T_\mathrm{c}$ based on the McMillian-Allen-Dynes formula suggests that $\lambda \approx 1$ is needed to reproduce the optimal $T_\mathrm{c} = 32$~K in Ba$_{1-x}$K$_x$BiO$_3$. Early theoretical estimates based on \gls*{DFT}~\cite{MeregalliPRB1998} instead placed $\lambda \approx 0.33$ (integrated over all branches), with the strongest coupling occurring to an optical phonon branch dispersing between $66-74$ meV. However, recent calculations using more sophisticated hybrid functionals found that the inclusion of the long-range Coulomb interaction could enhance the $e$-ph coupling, leading to $\lambda$ between 1 and 1.3 \cite{YinPRX2013}. Similarly, \gls*{DFT} calculations emphasizing the negative charge transfer character of the bismuthates also derive strong $e$-ph coupling to the $\mathbf{q}=(\pi,\pi,\pi)a^{-1}$ mode of the optical bond-stretching phonon branch and estimated  $\lambda\approx 0.89$ \cite{KhazraiePRB2018}. 

These results call for a re-evaluation of the $e$-ph interaction in these materials and its role in establishing the \gls*{CDW} phase at low doping. 
However, given the large $\lambda$ values estimated above, which are well outside the range where Migdal-Eliashberg theory is expected to be valid~\cite{EsterlisPRB2018}, it is crucial to perform such an analysis using numerically exact methods capable of treating possible polaronic effects.

Here we study the \gls*{CDW} transition in BaBiO$_3$, the parent compound of the bismuthate superconductors, using a scalable \gls*{HMC} algorithm \cite{CohenSteadPRE2022}. We consider a multi-orbital model that includes the Bi $6s$ and O $2p_\sigma$ orbitals and coupling to the Bi-O bond-stretching branch of optical phonons via the \gls*{SSH}-like (or Peierls-like) modulation of the Bi-O hopping integral. 
A highly efficient numerical code \cite{ElPhDynamics} makes possible simulations of large cubic clusters with input model parameters taken directly from  {\it ab initio} electronic structure calculations \cite{KhazraiePRB2018}, and a physically motivated phonon energy $\hbar\Omega_0 =  60$~meV \cite{BradenEPL1996, MeregalliPRB1998}. Our results demonstrate that an \gls*{SSH} coupling mechanism to the bond-stretching optical oxygen modes accounts for all aspects of the \gls*{CDW} phase in this compound without invoking any additional enhancement of the $e$-ph coupling constant beyond the early \gls*{DFT}-derived values once the polaronic effects are accounted for. Just as a complete understanding of the antiferromagnetic phase of the cuprates, provided by numerical solutions of the half-filled Hubbard Hamiltonian \cite{WhitePRB1989}, has been an essential foundation for the rich physics created by doping~\cite{ScalapinoRMP}, our work provides for a similar computationally rigorous solution of a model appropriate to the CDW phase of the bismuthates.

\section{Results}

\begin{figure*}
    \centering
    \includegraphics[width=\textwidth]{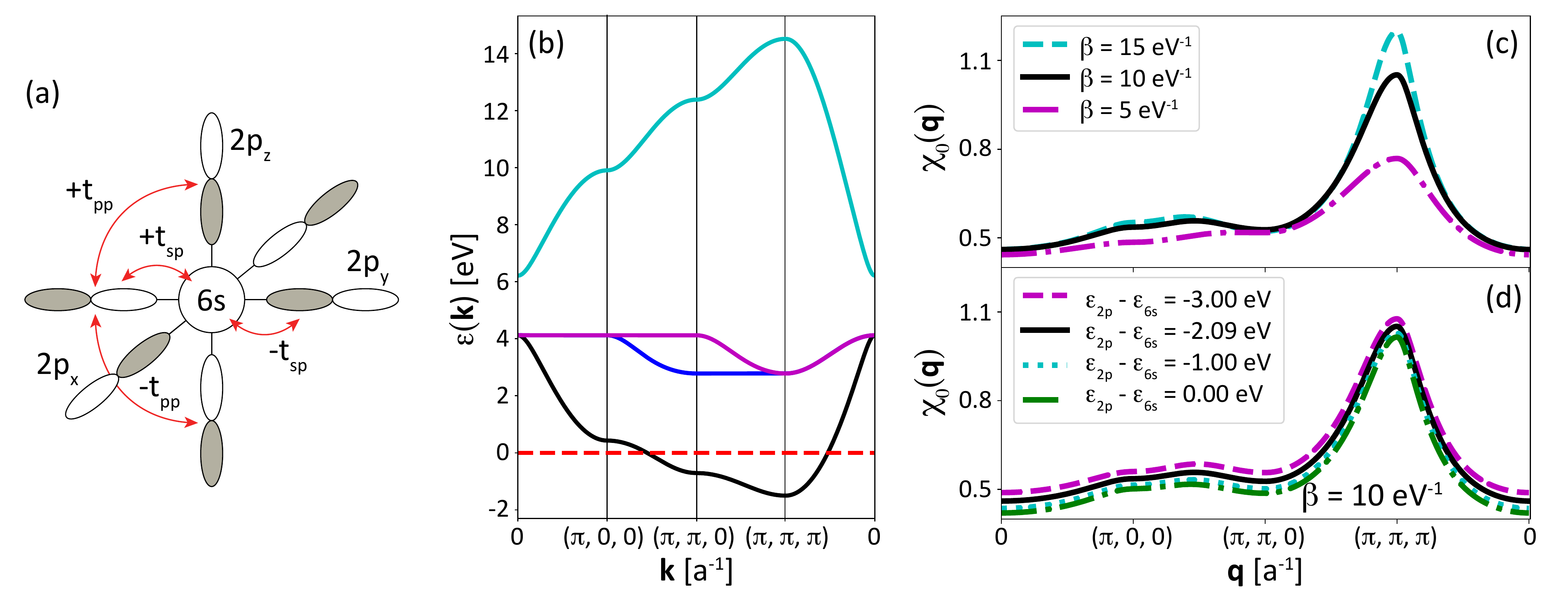}
    \caption{\textbf{Model definition and electronic properties in the noninteracting limit.} (\textbf{a}) Lattice structure of the three-dimensional, four-orbital model. A Bi $6s$ orbital is surrounding by six O $2p_\sigma$ orbitals. Red arrows indicate the relevant hopping integrals, with the phase factor included in hole language. (\textbf{b}) The band structure in the noninteracting limit, $\alpha_{sp}=0$. The horizontal red line indicates the Fermi energy at a target filling of 1 hole per Bi atom. (\textbf{c}) For the target filling, the noninteracting charge susceptibility $\chi_0(\mathbf{q})$ is computed using the lowest energy band (black) in panel (b) for three different temperatures. As the temperature is lowered, the peak at $\mathbf{q}=(\pi,\pi,\pi)a^{-1}$ grows. (\textbf{d}) At the fixed temperature $\beta = 10 \ \textrm{eV}^{-1}$, $\chi_0(\mathbf{q})$ for four values of the charge transfer energy $\epsilon_{2p}-\epsilon_{6s}$, the difference in the on-site energy of the O $2p_\sigma$ and Bi $6s$ orbitals. The peak at $\mathbf{q}=(\pi,\pi,\pi)a^{-1}$ is approximately independent of the charge transfer energy.}
    \label{fig:model}
\end{figure*} 

\subsection{The model}

We consider a four-orbital $sp$-model for BaBiO$_3$ defined on an $N = L^3$ cubic perovskite lattice (a total of $4L^3$ orbitals). The orbital basis includes the Bi~$6s$ and O~$2p_\sigma$ orbitals polarized along the Bi-Bi bond directions, as shown in Fig.~\ref{fig:model}(a). The Hamiltonian, written in hole language, is $\hat{H} = \hat{H}_e + \hat{H}_\textrm{ph} + \hat{H}_{e\textrm{-ph}}$, where
\begin{align}\label{eq:H0}
    \hat{H}_{e} =& (\epsilon_s-\mu)\sum_{\mathbf{i},\sigma} \hat{n}_{\mathbf{i},s,\sigma} + (\epsilon_p-\mu)\sum_{\mathbf{i},\vert \delta \vert, \sigma} \hat{n}_{\mathbf{i},\vert \delta \vert, \sigma}\\\nonumber
    &+ \sum_{\mathbf{i},\delta,\sigma} \mathcal{Q}^{\phantom{\dagger}}_{\delta} t^{\phantom{\dagger}}_{sp} [\hat{s}_{\mathbf{i},\sigma}^\dagger \hat{p}_{\mathbf{i},\delta,\sigma}^{\phantom{\dagger}} + \textrm{h.c.}]\\\nonumber
    &+ \sum_{\mathbf{i},\langle \delta, \delta' \rangle,\sigma} \mathcal{Q}^{\phantom{\dagger}}_{\delta,\delta'}t^{\phantom\dagger}_{pp}[\hat{p}_{\mathbf{i},\delta,\sigma}^\dagger \hat{p}_{\mathbf{i},\delta',\sigma}^{\phantom{\dagger}} + \textrm{h.c.}]
\end{align}
describes the noninteracting electron degrees of freedom, 
\begin{equation}
    \hat{H}_{\textrm{ph}} = \sum_{\mathbf{i},\vert \delta \vert} \bigg[ \frac{1}{2M}\hat{P}_{\mathbf{i},\vert \delta \vert}^2 + \frac{1}{2}M\Omega_0^2 \hat{X}_{\mathbf{i},\vert \delta \vert}^2 \bigg]
\end{equation}
describes the noninteracting lattice degrees of freedom, 
and
\begin{equation}
    \hat{H}_{e\textrm{-ph}} = \sum_{\mathbf{i},\delta,\sigma} \alpha^{\phantom{\dagger}}_{sp} \hat{X}^{\phantom{\dagger}}_{\mathbf{i},\delta} [\hat{s}_{\mathbf{i},\sigma}^\dagger \hat{p}_{\mathbf{i},\delta,\sigma}^{\phantom{\dagger}} + \textrm{h.c.}]
\end{equation}
describes the $e$-ph interaction. 
Here, $\hat{s}^\dagger_{\bf{i},\sigma}$ and $\hat{p}^\dagger_{\bf{i}, \vert \delta \vert,\sigma}$ create a spin-$\sigma$ hole on a Bi $6s$ and O $2p_\delta$ orbital in the unit cell containing a Bi atom located at $\mathbf{i}$ in the lattice, respectively. A sum over $\delta \ (\vert \delta \vert)$ runs over the set $\delta \in \{\pm x, \pm y, \pm z\} \ (\vert \delta \vert \in \{x, y, z\})$, where we have introduced the shorthand $\hat{p}_{\mathbf{i},-\vert \delta \vert, \sigma} = \hat{p}_{\mathbf{i}-\tfrac{a}{2}\vert\boldsymbol{\delta}\vert,\vert\delta \vert, \sigma}$ and $\hat{X}_{\mathbf{i},-|\delta|} = \hat{X}_{\mathbf{i}-\tfrac{a}{2}|\boldsymbol{\delta}|,|\delta|}$, such that $a$ is the nearest neighbor Bi-Bi distance. A sum over $\langle \delta, \delta' \rangle$ denotes a sum over nearest-neighbor O $2p$ orbital pairs that surround a central Bi $6s$ orbital. Finally, $\hat{n}^{\phantom{\dagger}}_{\mathbf{i},s,\sigma} = \hat{s}_{\mathbf{i},\sigma}^\dagger \hat{s}^{\phantom{\dagger}}_{\mathbf{i},\sigma}$ and $\hat{n}^{\phantom{\dagger}}_{\mathbf{i},\vert\delta\vert,\sigma} = \hat{p}_{\mathbf{i},\vert \delta\vert ,\sigma}^\dagger \hat{p}^{\phantom{\dagger}}_{\mathbf{i},\vert\delta\vert,\sigma}$ are the spin-$\sigma$ hole number operators for a Bi $6s$ and O $2p$ orbital in unit cell $\mathbf{i}$, respectively.

The on-site energies for the Bi $6s$ and O $2p$ orbitals are $\epsilon_s$ and $\epsilon_p$, respectively; $t_{sp}$ and $t_{pp}$ are the amplitude of the nearest neighbor Bi-O and O-O hopping integrals, respectively. The overall sign of the hopping integral is given by the phase factors $\mathcal{Q}_{\delta}$ and $\mathcal{Q}_{\delta,\delta'}$, such that $\mathcal{Q}_{\pm\delta}=\mp 1$, $\mathcal{Q}_{\pm \vert \delta \vert, \pm \vert \delta' \vert}=1$ 
and $\mathcal{Q}_{\pm \vert \delta \vert, \mp \vert \delta' \vert}=-1$, 
where $\delta\ne\delta'$. The sign convention enforced by these phase factors is shown in Fig.~\ref{fig:model}(a). $\hat{X}_{\mathbf{i},\vert \delta\vert}$ and $\hat{P}_{\mathbf{i},\vert\delta\vert}$ are the position and momentum operators for each oxygen atom, where $M$ is the mass of an oxygen atom, $\Omega_0$ is the bare phonon frequency, and $\alpha_{sp}$ is the $e$-ph coupling strength. 

Throughout, we use tight-binding parameters obtained from {\it ab initio} electronic structure calculations~\cite{KhazraiePRB2018}. Accordingly, we set the amplitude of the hopping integrals to $t_{sp} = 2.31$ and $t_{pp} = 0.335 \ \textrm{eV}$. The on-site energies are set to $\epsilon_s = 6.23$ and $\epsilon_p = 4.14 \ \textrm{eV}$ unless otherwise stated, consistent with the negative charge-transfer nature of these materials \cite{FoyevtsovaPRB2015, PlumbPRL2016, KhazraiePRB2018}. We adjust the chemical potential to obtain an average filling of $\langle n \rangle = 1$ hole/Bi using a recently developed $\mu$-tuning algorithm~\cite{MilesPRE2022}. Fig.~\ref{fig:model}(b) displays the corresponding hole band structure for this model with $\alpha_{sp}  = 0$. The horizontal dashed red line indicates the location of the Fermi energy at the target filling, which only intersects the lowest energy band. Based on this electronic structure, one might be tempted to adopt a single-band description of the system; however, we will demonstrate that the \gls*{SSH}-like interaction, which is off-diagonal in orbital space, necessitates a  multi-orbital description of the problem. 

To estimate the $e$-ph coupling $\alpha_{sp}$, we assume that the $sp$ hopping integrals depend on the oxygen displacement along the Bi-Bi bond direction $X$ according to $t_{sp}(X) = t_{sp} \big[1 + \tfrac{X}{a/2}\big]^{-2}$, where $a = 4.34 \ \textrm{\AA}$~\cite{HarrisonBook}. By expanding $t_{sp}(X)$ for small $X$
and assuming a linear coupling, we obtain $\alpha_{sp} = \frac{4t_{sp}}{a} = 2.13 \ \mathrm{eV/\AA}$. This value corresponds to a bare dimensionless coupling $\lambda = 0.18$ in the standard Migdal-Eliashberg formulation, where  
\begin{equation}\label{eq:lambda}
\lambda = \frac{2}{\hbar\Omega_0 N(0) N^2}
\sum_{{\bf k},{\bf q}} |g({\bf k},{\bf q})|^2
\delta(\epsilon_{\bf k}-\mu)\delta(\epsilon_{{\bf k}+{\bf q}}-\mu).  
\end{equation}
Here, $N(0)$ is the density of states per spin at the Fermi energy and $g({\bf k},{\bf q})$ is the $e$-ph coupling constant in momentum space. 
This value is consistent with early \gls*{DFT}-based estimates~ \cite{MeregalliPRB1998} \textit{without} correlation-driven  enhancements~\cite{YinPRX2013} (see Methods for further details).

We solve the model using a recently proposed set of algorithms~\cite{CohenSteadPRE2022}. Specifically, \gls*{HMC}~\cite{Duane1987, BeylPRB2018} and Fourier acceleration are used to efficiently sample decorrelated phonon fields~\cite{BatrouniPRB2019}. A physics-inspired preconditioner reduces the cost of each integration time-step. Efficient measurement of quantum observables is achieved using stochastic techniques that leverage the fast Fourier transform. See Methods
for more information. This approach allows us to simulate system sizes up to $L = 10$ ($4 L^3 = 4000$ orbitals) and consider optical phonons with energies comparable to those in the real system. To this end, we take $\hbar\Omega_0 = 60$ meV, which is typical for the bond-stretching optical oxygen phonons in most oxides~\cite{KhosroabadiPRB2011, MeregalliPRB1998}. Note that in all of our simulations, we have confirmed that the displacements of the oxygen atoms remain small enough that the effective hopping never changes sign, ensuring the validity of a linear \gls*{SSH} coupling ~\cite{Prokofev22, LiPreprint2022}.

To detect charge order we measure the static charge structure factor $S_{\gamma\textrm{-}\nu}(\mathbf{q}) = S_{\gamma\textrm{-}\nu}(\mathbf{q}, \tau=0)$ given by
\begin{align}\label{eq:phonon_dispersion}
    S_{\gamma\textrm{-}\nu}(\mathbf{q},\tau) = \frac{1}{N} \sum_{\mathbf{r},\mathbf{i}} e^{-\rm{i}\mathbf{q}\cdot\mathbf{r}} \langle \hat{n}_{\mathbf{i}+\mathbf{r},\gamma}(\tau) \hat{n}_{\mathbf{i},\nu}(0) \rangle,
\end{align}
and the corresponding charge susceptibility
\begin{align}
    \chi_{\gamma\textrm{-}\nu}(\mathbf{q}) = \int_0^\beta S_{\gamma\textrm{-}\nu}(\mathbf{q}, \tau) \ d\tau,
\end{align}
where $\hat{n}_{\mathbf{i},\nu} = (\hat{n}_{\mathbf{i},\nu,\uparrow} + \hat{n}_{\mathbf{i},\nu,\downarrow})$, and $\gamma$ and $\nu$ specify the orbital species.

We also calculate the renormalized phonon energy
\begin{align}\label{eq:omega_dressed}
\hbar\Omega(\mathbf{q},0) = \sqrt{\hbar^2\Omega_0^2+\Pi(\mathbf{q},0)},
\end{align}
where $\Pi(\mathbf{q},\nu_n)$ is a function related to the phonon self-energy and $\nu_n=2\pi n T$ is a bosonic Matsubara frequency. 
We obtain $\hbar\Omega(\mathbf{q},0)$ by measuring the phonon Green's function 
\begin{align}
    D(\mathbf{r},\tau) = -\frac{2M\Omega_0}{3N\hbar} \sum_{\mathbf{i},\vert\delta\vert} \langle \hat{X}_{\mathbf{r}+\mathbf{i},\vert \delta\vert}(\tau) \hat{X}_{\mathbf{i},\vert\delta\vert}(0) \rangle,
\end{align}
where the Fourier transform in both $\mathbf{r}$ and $\tau$ is related to $\Pi({\bf q},\nu_n)$ by \cite{EsterlisPRB2018}
\begin{equation}
    D(\mathbf{q},\rm{i}\nu_n) = \frac{2\hbar\Omega_0}{(\rm{i}\nu_n)^2-\hbar^2\Omega_0^2-\Pi(\mathbf{q},\rm{i}\nu_n)}.
\end{equation}

\begin{figure}
    \centering
    \includegraphics[width=\columnwidth]{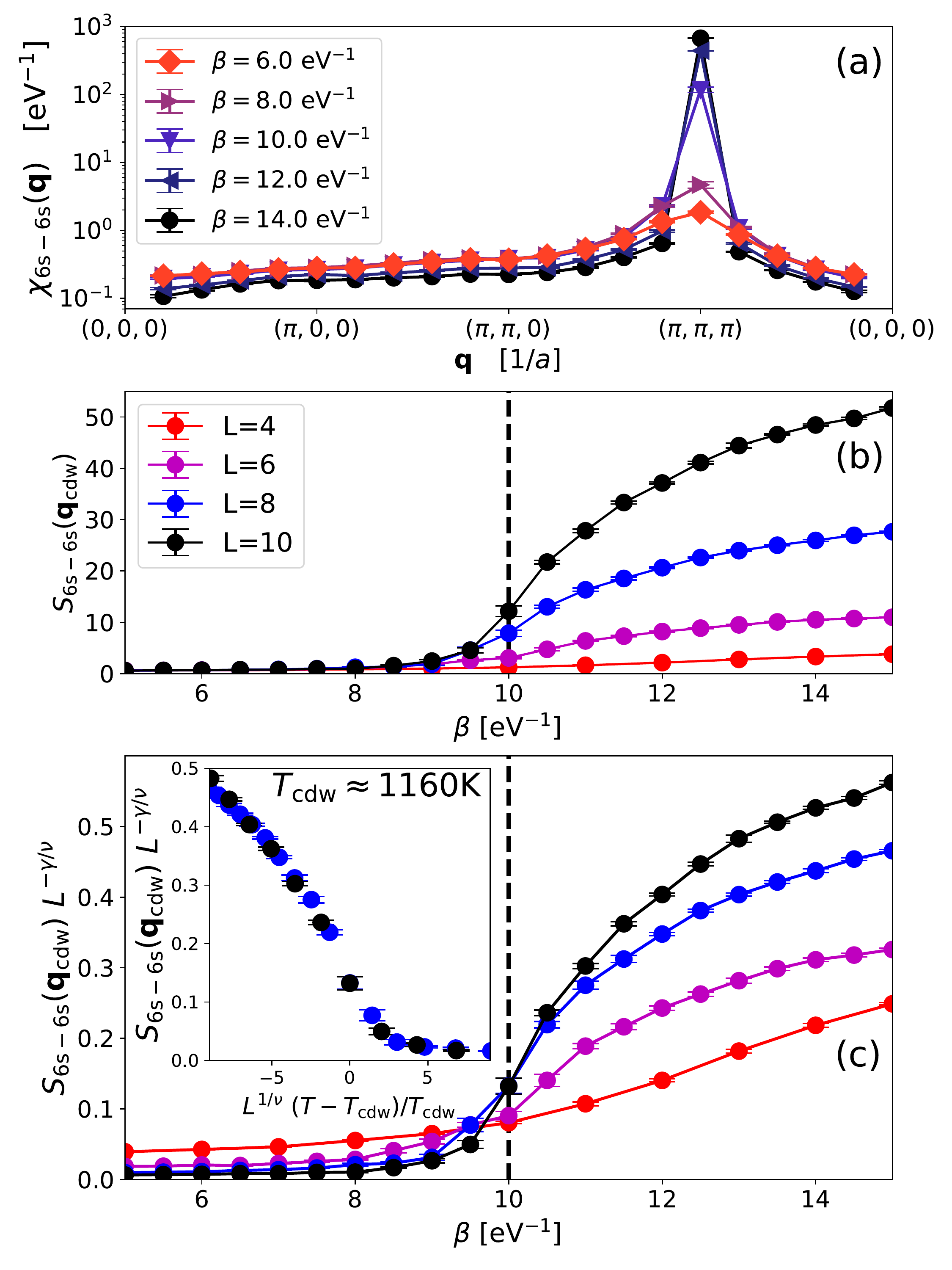}
    \caption{\textbf{Finite size scaling to determine transition temperature }$T_\textrm{cdw}$. (\textbf{a}) The charge susceptibility $\chi_\textrm{6s-6s}(\mathbf{q})$ versus $\mathbf{q}$ for various $\beta$, plotted on a semilog scale for an $L=10$ cluster. As the temperature is lowered, a peak forms at $\mathbf{q}=(\pi,\pi,\pi)a^{-1}$. 
    (\textbf{b}) The charge structure factor $S_\mathrm{6s-6s}(\mathbf{q}_\textrm{cdw})$, sensitive to CDW order on Bi-6s orbitals, versus $\beta$. (\textbf{c}) Crossing plot where the vertical axis is scaled using the three-dimensional Ising critical exponents. The vertical dashed black line indicates the transition temperature $T_\textrm{cdw}=1160 \ K$. The inset displays the corresponding collapse of the charge structure factor for the $L = 8$ and $10$ data.
    The error bars in this figure are the standard deviation of the mean of our measurements.}
    \label{fig:fss}
\end{figure}

\subsection{Charge-density-wave phase transition in BaBiO$_3$}

Both formal valence counting and electronic structure calculations~ \cite{MeregalliPRB1998, LiechtensteinPRB1991, FoyevtsovaPRB2015} indicate that BaBiO$_3$, in its high-temperature metallic phase, has on average one hole/Bi. In this case, the large $t_{sp}$ hopping integral results in a half-filled $sp$ hybridized antibonding band (in electron language), which disperses through the Fermi level, as shown in Fig.~\ref{fig:model}(b) (hole language). Throughout this paper, we focus on this case and fix $\langle n \rangle = 1$ hole/Bi atom. 

According to high-resolution neutron powder diffraction measurements, 
BaBiO$_3$ is a \gls*{CDW} insulator over a wide temperature range ($4 - 973$~K)~\cite{Kennedy2006}, with the upper bound approaching the material's melting point \cite{KLINKOVA1999439}. 
The insulating phase is characterized by a $\mathbf{q}_\textrm{cdw}=(\pi,\pi,\pi)/a$ ordering vector and expansion and contraction of alternating BiO$_6$ octahedra. This ordering leads to a double perovskite structure, with two inequivalent Bi sites whose average Bi-O bond lengths are largely independent of temperature. We, therefore, begin by studying the \gls*{CDW} transition in our model. In all of our calculations, we find that the O-O correlations remain weak at all temperatures, so no charge modulations develop on the oxygen sites at any temperature. This result is consistent with a bond-disproportionation scenario, where all oxygen orbitals are equivalent at all temperatures \cite{FoyevtsovaPRB2015, KhazraiePRB2018}. 
Looking at the dressed charge susceptibility on the Bi sites $\chi_\textrm{6s-6s}(\mathbf{q})$ versus $\mathbf{q}$, as shown in Fig.~\ref{fig:fss}(a), we see a narrow peak form at the expected ordering wave-vector $\mathbf{q}_\textrm{cdw}$. 
Fig.~\ref{fig:fss}(b), therefore, focuses on the Bi charge structure factor $S_\textrm{6s-6s}(\mathbf{q}_\textrm{cdw})$, which reflects the formation of a charge modulation 
in which there are distinct local densities on two sublattices of the Bi sites.
We find that $S_\textrm{6s-6s}(\mathbf{q}_\textrm{cdw})$ increases with decreasing temperature, with the low-temperature structure factor also growing with system size $L$ at lower temperatures. This behavior is indicative of the formation of long-range order. 

As this \gls*{CDW} phase spontaneously breaks a $\mathbb{Z}_2$ symmetry between the two otherwise equivalent Bi $6s$ sub-lattices, this phase transition falls in the three-dimensional Ising universality class. Therefore, we use the corresponding critical exponents $\gamma = 1.237$ and $\nu=0.630$ to perform a finite-size scaling (FSS) analysis to determine the transition temperature \cite{Hasenbusch2010}. Figure~\ref{fig:fss}(c) displays the relevant  $S_\textrm{6s-6s}(\mathbf{q}_\textrm{cdw}) L^{-\gamma/\nu}$ versus $\beta$ curves. Here, the locus of crossing points of the data for each cluster size provides an estimate of the critical temperature, provided the clusters are large enough to capture the relevant spatial correlations. Using this data for the crossing of the $L = 8$ and $10$ data sets, we obtain $T_\textrm{cdw}\approx 10~\mathrm{eV}^{-1} = 1160$ K. This value is above the material's melting point \cite{KLINKOVA1999439} and thus explains why this material remains a \gls*{CDW} insulator at all temperatures. The inset panel displays the corresponding collapse that occurs when $S_\textrm{6s-6s}(\mathbf{q}_\textrm{cdw}) L^{-\gamma/\nu}$ is plotted against $L^{1/\nu}(T-T_\textrm{cdw})/T_\textrm{cdw}$, demonstrating the  quality of the scaling analysis. In this case, we have omitted the $L=4$ and $6$ results from the collapse. Note that a similar analysis for the \gls*{CDW} transition in the 3D Holstein model also found that $L=4$ and $6$ clusters lie outside the scaling regime and need to be excluded from a finite size scaling analysis~\cite{CohenSteadPRB2020}.  \\

\begin{figure}
    \centering
    \includegraphics[width=\columnwidth]{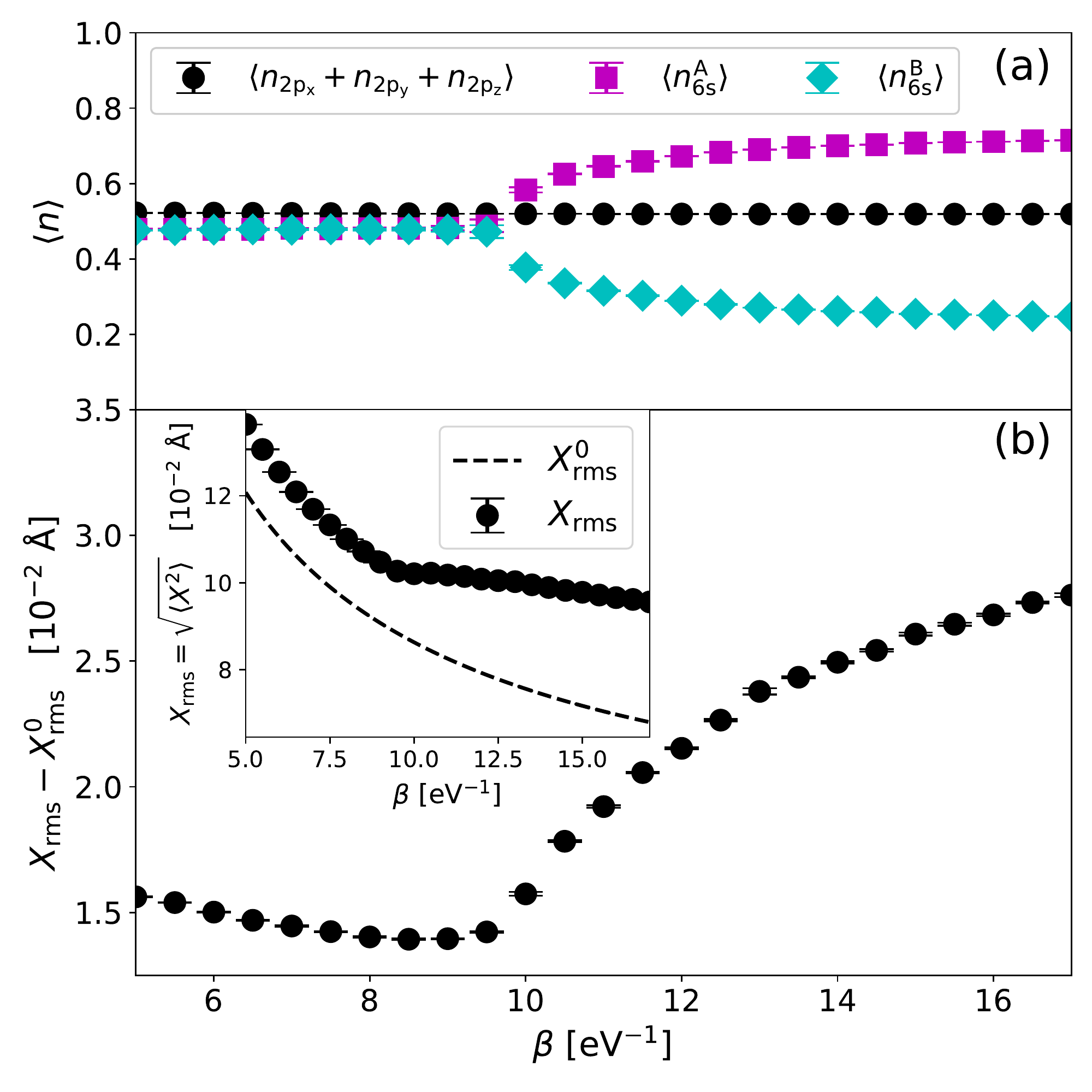}
    \caption{{\bf Local quantities versus $\beta$ for the $L=10$ system.} ({\bf a}) The total density of holes on the O $2p_\sigma$ orbitals (black circles) remains approximately independent of $\beta$ as the temperature is lowered below $T_\mathrm{cdw}$. The hole densities on the Bi $6s$ orbital $A$ and $B$ sub-lattices, $\langle n_\textrm{6s}^\textrm{A}\rangle$ and $\langle n_\textrm{6s}^\textrm{B}\rangle$ respectively, bifurcate near $T_\mathrm{cdw}$, indicating the onset of charge order. ({\bf b}) The inset shows the root mean square fluctuations in the phonon position as measured in the presence (black circles) and the absence of interactions, $\alpha_{sp}=0$ (dashed line). The main figure displays the difference between these two quantities, with an inflection point occurring near $T_\textrm{cdw}$.
    The error bars in this figure are the standard deviation of the mean of our measurements.}
    \label{fig:local}
\end{figure}

\subsection{Local quantities}

A signature of the phase transition is also present in the local densities shown in Fig.~\ref{fig:local}(a). At temperatures above the transition, $T > T_\textrm{cdw}$, the average hole density on the Bi $6s$ sub-lattices are equal, $\langle n_{6s}^\textrm{A} \rangle \approx \langle n_{6s}^\textrm{B} \rangle$. 
For $T < T_\textrm{cdw}$ these two local densities bifurcate with the onset of \gls*{CDW} order. Notably, the hole density on the oxygen atoms remains approximately constant across the transition. In other words, there is no net transfer of charge to the ligand oxygen atoms as charge modulations form on the Bi $6s$ orbitals. This behavior is reminiscent of the behavior seen in models for the rare-earth nickelates $R$NiO$_3$ \cite{ParkPRL2012, JohnstonPRL2014}, which are also negative charge transfer systems.

At low temperatures the difference in the hole density on each of the two Bi-$6s$ sub-lattices begins to saturate, with $\langle n_{6s}^\textrm{A} \rangle - \langle n_{6s}^\textrm{B} \rangle \approx 0.47$ at $\beta = 17 \ \textrm{eV}^{-1}$. It has been proposed that the formation of \gls*{CDW} order in BaBiO$_3$ is driven by charge disproportionation 
and the formation of a mixed valence state $\textrm{Bi}^{4+\delta}\textrm{Bi}^{4-\delta}$, corresponding to $\langle n_{6s}^\textrm{A} \rangle - \langle n_{6s}^\textrm{B}\rangle = 2\delta$. In the extreme case, the charge disproportionation is complete, with $\delta = 1$. Our model, however, predicts a much smaller charge transfer between neighboring Bi orbitals, which is more in line with experimental observations~ \cite{WertheimPRB1982, PlumbPRL2016}.

These results indicate that the holes in the compressed octahedra are not localized on a single Bi atom but instead occupy a more spatially distributed state involving the ligand oxygen orbitals. Previous investigations suggest this state is well described by hybridization between a Bi $6s$ orbital and an $A_\mathrm{1g}$ molecular orbital formed from the six surrounding O $2p_\sigma$ orbitals~\cite{FoyevtsovaPRB2015, LiQM2020}. The formation of this spatially extended hybridized state coincides with the oxygen octahedron surrounding a central bismuth atom contracting, thus forming a bipolaron. Conversely, the oxygen octahedron surrounding the six
adjacent bismuth atoms must expand. There is a clear signature of this behavior in Fig.~\ref{fig:local}(b). The inset shows both the measured root mean square of the fluctuations in the phonon position, $X_\textrm{rms}=\sqrt{\langle X^2 \rangle}$, as well as the analytic value $X^0_\textrm{rms}$ in the $\alpha_{sp}=0 \ {\rm eV / \AA}$ limit, with the main panel displaying the difference $X^{\phantom{0}}_\textrm{rms} - X^0_\textrm{rms}$. For $T>T_\textrm{cdw}$, this difference initially decreases as the temperature is lowered, reflecting the reduction of thermal fluctuations of the lattice. However, at $T \approx T_\textrm{cdw}$ it increases rapidly, then gradually levels off as the temperature is lowered further. This behavior can be attributed to the formation of an alternating pattern of expanded and contracted oxygen octahedron associated with the onset of \gls*{CDW} order.
At temperatures well into the \gls*{CDW} phase, we obtain an $X_\textrm{rms}$ that is nearly temperature independent and consistent with the experimental values obtained from high-resolution neutron powder diffraction. For example, at $\beta = 13$ eV$^{-1}$ (893 K), we obtained $X_\mathrm{rms} = 0.100 \pm 0.008$~\AA, 
which agrees with the experimental value $0.094~\mathrm{\AA}$ measured at this temperature~\cite{Kennedy2006}. 
These results explain why the measured Bi-O bond lengths in BaBiO$_3$ are essentially independent of temperature in the \gls*{CDW} insulating phase~\cite{Kennedy2006}. They also 
confirm that our model yields physically reasonable displacements and that no unphysical sign changes occur in the effective hopping integrals $t_{sp}\pm\alpha_{sp}\langle X\rangle$. \\

\begin{figure}
    \centering
    \includegraphics[width=\columnwidth]{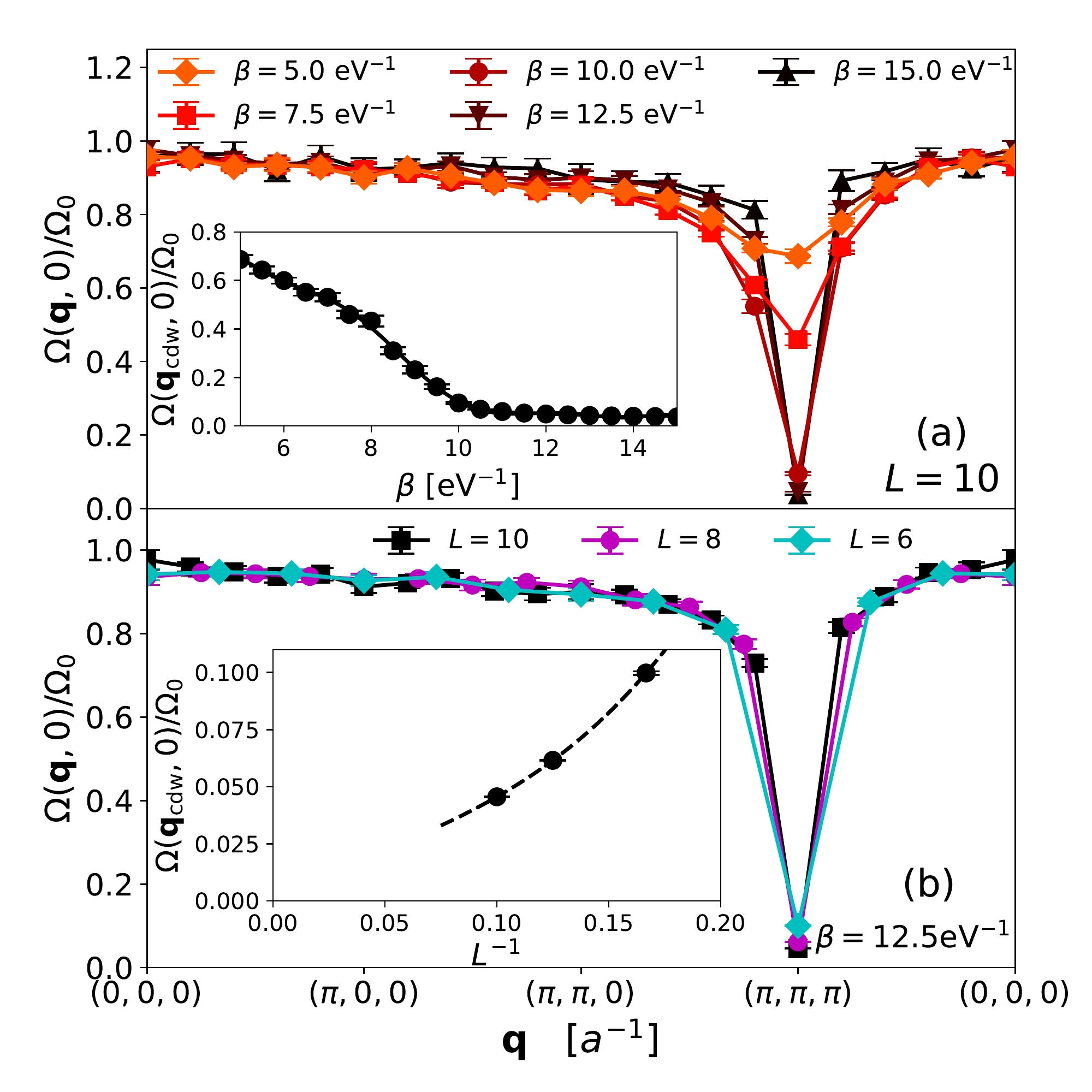}
    \caption{{\bf The renormalized phonon dispersion.}
    (\textbf{a}) $\Omega(\mathbf{q},\mathrm{i}\nu_n=0)$ along the high-symmetry cuts of the first Brillouin zone, computed on an $L = 10$ lattice. 
    As the temperature is lowered $\Omega(\mathbf{q},\mathrm{i}\nu_n=0)$ softens at the ordering wave vector $\mathbf{q}=(\pi,\pi,\pi)a^{-1}$, with other $\mathbf{q}$ remaining essentially pinned at $\Omega_0$. The inset shows the softening of the $\Omega(\mathbf{q}_\textrm{cdw},0)$ mode specifically versus $\beta$. (\textbf{b}) $\Omega(\mathbf{q},\mathrm{i}\nu_n=0)$ for different lattice sizes, measured at a fixed inverse temperature $\beta = 12.5~{\rm eV}^{-1}$ below the transition temperature. Here,  $\Omega(\mathbf{q},0)$ softens at $\mathbf{q}_{\rm cdw}$ for all lattice sizes $L = 6,8, \textrm{ and } 10$. The inset shows $\Omega(\mathbf{q}_{\rm cdw},0)$ decreasing with increasing lattice size, with the dashed line acting as a guide for the eye. The error bars in this figure are the standard deviation of the mean of our measurements.}
    \label{fig:phonon_branch}
\end{figure}

\subsection{Phonon renormalizations}

To gain further insight into the nature of the \gls*{CDW} transition, we calculated the renormalized phonon energy $\Omega(\mathbf{q},0)$ as a function of scattering momentum $\mathbf{q}$. In the $\alpha_{sp}=0 \ \textrm{eV/\AA}$ limit, the vibrations of the oxygen atoms in the Bi-Bi bond direction are described by a dispersionless optical phonon branch $\Omega(\mathbf{q},0) = \Omega_0$. Figure~\ref{fig:phonon_branch} shows that this behavior is approximately recovered at  temperatures well above $T_\textrm{cdw}$, with only a relatively small dip in $\Omega(\mathbf{q},0)$ occurring at the ordering wave-vector $\mathbf{q}_\textrm{cdw}$. 
As the temperature is lowered, however,  $\Omega(\mathbf{q},0)$ is significantly renormalized, with a pronounced softening of the phonon branch occurring at $\mathbf{q}_\textrm{cdw}$. 
Moreover, this softening is quite sharp; at $\beta = 15.0 \ \textrm{eV}^{-1}$ we see that $\Omega(\mathbf{q}\ne\mathbf{q}_\textrm{cdw},0) \approx \Omega_0$, while $\Omega(\mathbf{q}_\textrm{cdw},0)$ has softened to a value close to zero. The inset of Fig.~\ref{fig:phonon_branch}(a)
shows that for $T < T_\textrm{cdw}$, the frequency of the  $\mathbf{q}_\textrm{cdw}$ phonon mode levels off at a fixed value close to zero. The fact that it does not reach zero exactly is an expected finite size effect [see Fig.~\ref{fig:phonon_branch}(b)],
with a true divergence in $\chi({\bf  q}_\mathrm{cdw})$ and complete softening only expected in the thermodynamic limit.

Our results for the average Bi $6s$ occupations and Bi-O bond distances demonstrate that the \gls*{CDW} transition in BaBiO$_3$ can be captured using the bond-disproportionation picture associated with the negative charge transfer nature of its electronic structure. However, Ref.~\cite{KhazraiePRB2018b} has pointed out that the charge and bond-disproportionation are just two ends of a continuous spectrum of possibilities. With this in mind, we next consider what effect modifying the on-site energy difference $\Delta \epsilon = (\epsilon_p - \epsilon_s)$ has on the character of the \gls*{CDW} order. This analysis will also reveal some subtle aspects of this transition. \\

\begin{figure}
    \centering
    \includegraphics[width=\columnwidth]{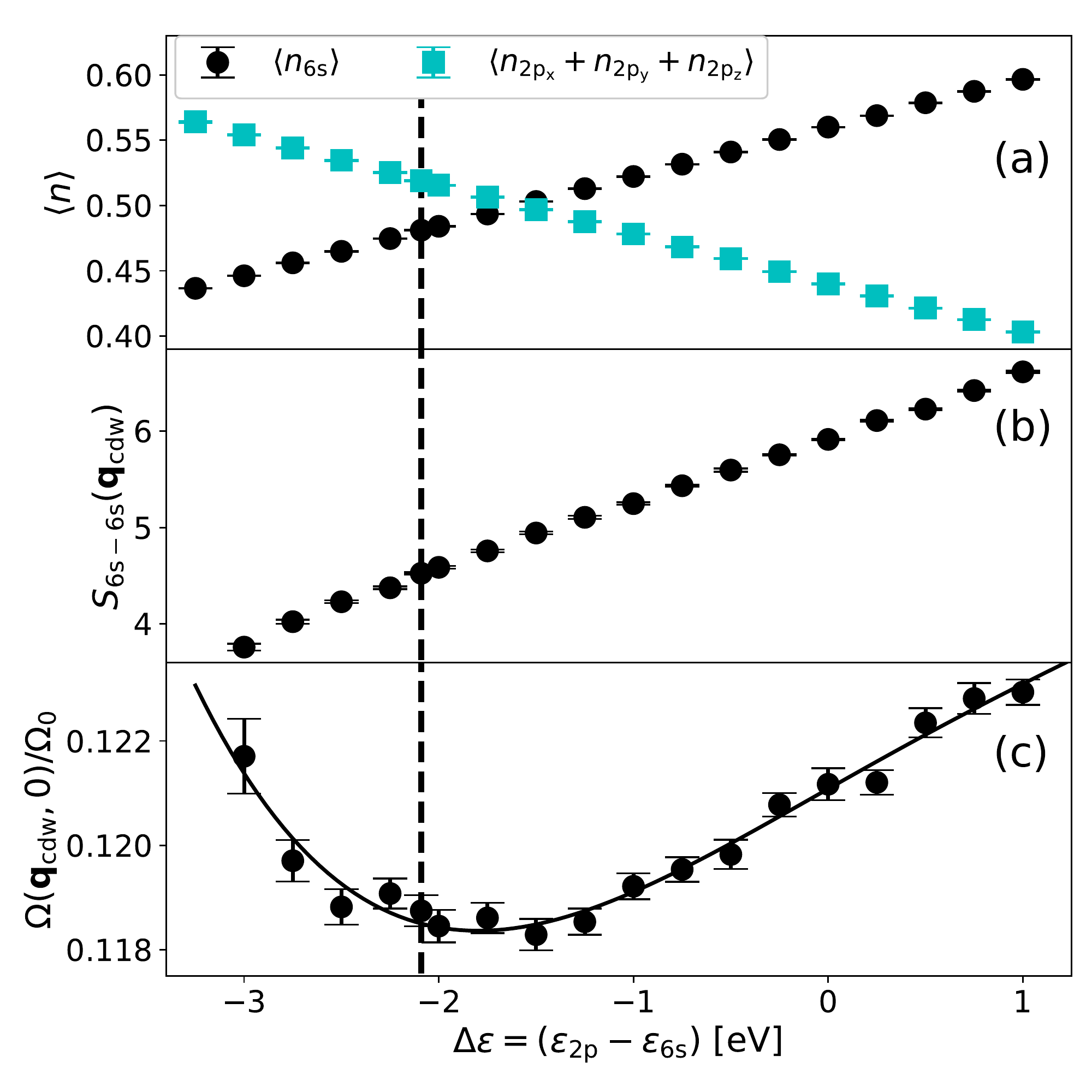}
    \caption{{\bf Sweep through on-site energy difference $\Delta\epsilon=(\epsilon_p-\epsilon_s)$.} Results were collected on an $L=4$ lattice at inverse temperature $\beta=18 \ \textrm{eV}^{-1}$. The vertical dashed lines indicates the \textit{ab initio} value of $\Delta\epsilon$. (\textbf{a}) The total hole density on the O $2p_\sigma$ orbitals (cyan) and the Bi $6s$ orbitals versus $\Delta\epsilon$. Near $\Delta\epsilon \approx -1.7$ eV, these two hole densities are commensurate. (\textbf{b}) The structure factor $S_\textrm{6s-6s}(\mathbf{q}_\textrm{cdw})$ monotonically increases with $\Delta\epsilon$. (\textbf{c}) The renormalized phonon energy of the $\Omega(\mathbf{q}_\textrm{cdw},0)$ mode has a local minimum near $\Delta\epsilon \approx -1.7$ eV. The black line is a guide for the eye. 
    The error bars in this figure are the standard deviation of the mean of our measurements.}
    \label{fig:energy_sweep}
\end{figure}

\subsection{The role of the charge-transfer energy}

Figure~\ref{fig:energy_sweep}(a) shows how the average density of holes on the oxygen and bismuth atoms changes with $\Delta\epsilon$. In particular, for $\Delta\epsilon \approx -1.7$ eV, the average total density of holes on the oxygen atoms equals the hole density on bismuth atoms, $\langle n_{6s} \rangle \approx \langle n_{2p_x} + n_{2p_y} + n_{2p_z} \rangle$. Increasing (decreasing) $\Delta\epsilon$ transfers holes to the bismuth (oxygen) orbitals. 

Figure~\ref{fig:energy_sweep}(b) examines how the distribution of holes affects the charge correlations. Increasing $\Delta\epsilon$ not only increases $\langle n_{6s} \rangle$ but  $S_\textrm{6s-6s}(\mathbf{q}_\textrm{cdw})$ as well. 
Referring back to Figs.~\ref{fig:model}(c-d), we see that the non-interacting charge susceptibility $\chi_0(\mathbf{q})$ has a peak at $\mathbf{q}_{\textrm{cdw}}$ that grows with $\beta$. However, the height of this peak also decreases monotonically with increasing $\Delta\epsilon$. We can therefore conclude that the growth in $S_\textrm{6s-6s}(\mathbf{q}_\textrm{cdw})$ with $\Delta\epsilon$ is not the result of changes in the bare electronic structure and the degree of  Fermi surface nesting. Rather, the increase in $S_\textrm{6s-6s}(\mathbf{q}_\textrm{cdw})$ can strictly be understood as resulting from the increasing hole density on the Bi-$6s$ orbitals. 

At face value, the observed increase in $S_\textrm{6s-6s}(\mathbf{q}_\textrm{cdw})$ is not surprising as a large Bi hole density can support a stronger charge modulation on these orbitals. What is surprising is that $S_\textrm{6s-6s}(\mathbf{q}_\textrm{cdw})$ does not appear to correlate to the degree of phonon softening strictly. For example, if you were to only look at  $S_\textrm{6s-6s}(\mathbf{q}_\textrm{cdw})$,
you might conclude that the \gls*{CDW} correlations grow by nearly 33\% as $\Delta \epsilon$ is changed from $-3~\mathrm{eV}$ to $1~\mathrm{eV}$. 
However, the \gls*{CDW} phase is characterized by both the hole density modulations and the periodic disproportionation of the Bi-O bond distances. As previously discussed, this periodic displacement of the oxygen atoms away from the midpoint between two neighboring bismuth atoms softens the $\mathbf{q}_\textrm{cdw}$ mode. In Fig.~\ref{fig:energy_sweep}(c) we see that $\Omega(\mathbf{q}_\textrm{cdw},0)$ is a nonmonotonic function of $\Delta\epsilon$ that does not track the dependence of $S_\textrm{6s-6s}(\mathbf{q}_\textrm{cdw})$ on the charge transfer energy. 
Moreover, $\Omega(\mathbf{q}_\textrm{cdw},0)$ is minimized near $\Delta\epsilon \approx -1.7$ eV, the same point at which $\langle n_{6s} \rangle \approx \langle n_{2p_x} + n_{2p_y} + n_{2p_z} \rangle$. This observation highlights the crucial role of the hybridization between the Bi $6s$ and surrounding O $2p_\sigma$ orbitals in establishing the \gls*{CDW} order. \\

\begin{figure}
    \centering
    \includegraphics[width=\columnwidth]{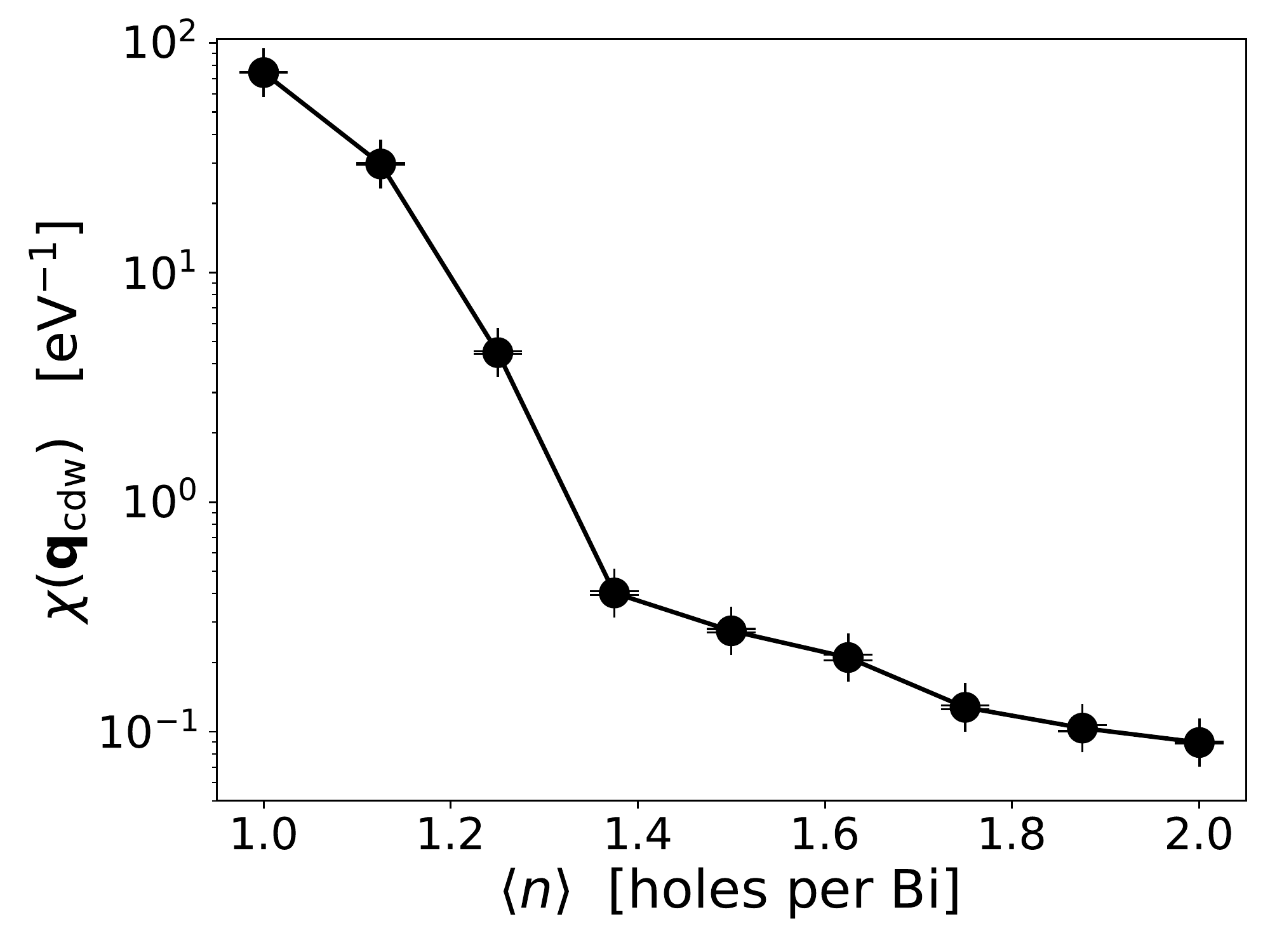}
    \caption{{\bf Change in susceptibilities upon doping.} Results were collected on an $L=4$ lattice at inverse temperature $\beta = 18~\textrm{eV}^{-1}$. The charge susceptibility $\chi_\textrm{6s,6s}(\mathbf{q}_\textrm{cdw})$ monotonically decreases upon doping. 
    The error bars in this figure are the standard deviation of the mean of our measurements.}
    \label{fig:doped_susceptibility}
\end{figure}

\subsection{Doping dependence of the charge correlations}

Our results demonstrate that coupling to the bond-stretching motion of the oxygen atoms in BaBiO$_3$ is 
sufficient to explain the \gls*{CDW} phase transition in this system at half-filling. 
To assess the relevance of this interaction for the doped system, we also calculated the 
charge susceptibility $\chi_\mathrm{6s-6s}({\bf q}_\mathrm{cdw})$   
as a function of doping, as shown in Fig.~\ref{fig:doped_susceptibility}. Here, we focus on an $L = 4$ cluster and an 
inverse temperature of $\beta = 18~\mathrm{eV}^{-1}$ ($\approx 644~\mathrm{K}$). 
While the strength of the 
\gls*{CDW} correlations decreases with doping, the correlations remain strong to carrier concentrations as large 
as $\langle n\rangle = 1.25$ holes/Bi. The strength of these correlations is consistent with the Ba$_{1-x}$K$_x$BiO$_3$ phase diagram, where the \gls*{CDW} phase extends out to $\langle n \rangle \approx  1.2$ holes/Bi at this temperature~\cite{Sleight2015}. These results further support the notion that the \gls*{SSH} interaction with a relatively weak bare dimensionless coupling $\lambda$ can account for many aspects of the bismuthate temperature-doping phase diagram. 

\section{Discussion}
Our model 
has a relatively weak bare dimensionless coupling $\lambda = 0.18$ when the $e$-ph coupling is characterized in the standard way [see Eq.~\eqref{eq:lambda} and Methods].
Moreover, the prominent peak in the bare charge susceptibility $\chi_0({\bf q})$ shown in Fig.~\ref{fig:model}(c) indicates that Fermi surface for the noninteracting band structure is well nested for the ordering wave-vector ${\bf  q}_\mathrm{cdw}$. Given this strong nesting condition, weak $e$-ph coupling, and the pronounced softening of the phonon dispersion predicted in Fig.~\ref{fig:phonon_branch}, one might be tempted to conclude that the \gls*{CDW} transition is driven by a  weak-coupling Peierls-like scenario. This picture, however, is inconsistent with many of our results. For example, a na{\"i}ve RPA calculation over predicts $T_\mathrm{cdw}$ by a factor of two (see Methods).
This discrepancy suggests that electron and phonon self-energies must be included to describe the \gls*{CDW} transition in this system. Such effects would also result in an effective coupling $\lambda_\mathrm{eff}$ that is much larger than the bare coupling computed from the noninteracting phononic and electronic structure.  In fact, if we use the renormalized phonon energies to calculate the effective total and mode-resolved dimensionless $e$-ph couplings (see Methods), we obtain $\lambda^\text{eff} = 0.22$ and $\lambda^\text{eff}_{{\bf q}_\text{cdw}} \approx 9.92$, respectively, at $\beta = 17$ eV$^{-1}$.
But what is perhaps more important is the fact that  
the RPA cannot account for the non-monotonic dependence of the soft phonon mode energy as a function of the $6s$-$2p$ (Bi-O) site energy difference $\Delta\epsilon$. Our \gls*{HMC} results show that $\Omega({\bf q}_\mathrm{cdw})$ is minimized when the hybridization between the Bi $6s$ and O $2p$ orbitals is the largest, which is a  consequence of the off-diagonal character of the \gls*{SSH} coupling and the {\it multi-orbital} nature of the problem. In contrast, a single-band RPA description predicts a monotonic behavior as a function of $\Delta\epsilon$, contrary to the exact numerics.  

Instead, our results point towards a bipolaronic scenario. In this picture, the ligand $A_\mathrm{1g}$ molecular orbital strongly hybridizes with the Bi $6s$ orbital at its center with an effective hopping $t_\mathrm{eff} = \sqrt{6}t_{sp}$ \cite{KhazraiePRB2018}. This hybridization 
creates bonding and antibonding states that shift $\pm t_\mathrm{eff}$ above and below the Fermi energy in hole language. A subsequent breathing distortion of the lattice creates a two sublattice structure and increases $t_\mathrm{eff}$. Therefore, the pair of holes originally residing on opposite sublattices in the undistorted structure can lower their kinetic energies significantly by condensing into the hybridized state below the Fermi energy. In this picture, the condensed holes and corresponding breathing distortions should be viewed as a bipolaron, and the \gls*{CDW} state as a frozen bipolaron lattice~\cite{LiQM2020}. 
This scenario naturally explains why the largest softening of the $\Omega({\bf q}_\mathrm{cdw})$ mode occurs at the point of maximal hybridization between the Bi $6s$ and O $2p$ orbitals since this state would also produce the largest shift in the antibonding states and, subsequently, the largest binding energy between the holes. \gls*{DFT} calculations have found that such a polaronic mechanism results in a significant coupling ($\approx 10~\mathrm{eV}/\mathrm{\AA}$) between the $A_\mathrm{1g}$ holes and the breathing lattice distortion~
\cite{KhazraiePRB2018}, which far exceeds the bare $e$-ph coupling strength estimated using more conventional DFT methods. This is consistent with our estimated enhancement of the mode-resolved coupling with 
$\lambda^\text{eff}_{{\bf q}_\mathrm{cdw}}/\lambda_{{\bf q}_\mathrm{cdw}} \approx 31 \approx \Omega_0/\Omega({{\bf q}_\mathrm{cdw}},0)$.

We note that this bipolaron scenario was also advocated for in Ref.~\onlinecite{LiQM2020} based on determinant quantum Monte Carlo simulations of small $4\times 4$ clusters in 2D, and for an unphysically large phonon energy. Our results demonstrate that this picture survives for the 3D lattice and a more realistic material-specific model. 
The bipolaronic description of the \gls*{CDW} phase holds for this system, despite the relatively small value of the bare dimensionless $e$-ph coupling $\lambda = 0.18$, which is below the (bare)
value $\lambda_c = 0.5$ where polaronic effects become significant in the 2D
Holstein model~\cite{EsterlisPRB2018}. These results thus underscore the notion that intuition derived from Holstein- and Fr{\"o}hlich-like models, which are diagonal in orbital space, may not serve us well when studying off-diagonal $e$-ph interactions. 

Our work also demonstrates that the early estimate for the bare coupling~\cite{MeregalliPRB1998} is sufficient to account for the \gls*{CDW} phase in the bismuthates and, therefore, calls for a re-evaluation of $e$-ph interactions in the doped metallic phase.   
Previous attempts to explore superconductivity mediated by CDW exchange within the random phase approximation (RPA)~\cite{RicePRL1981, bickers1987cdw} 
have examined an effective extended
Hubbard Hamiltonian in which an intersite electron-electron repulsion replaces the electron-phonon interaction
$V \sim \alpha^2/(\hbar\Omega_0)^2$ obtained by integrating out the phonons. While the resulting $s$-wave pairing interaction
strength and spectral weight could account for superconductivity at high temperatures, the calculations were
always considered problematic given the tendency of the RPA to overestimate $T_{\rm c}$ and also the absence of retardation
with instantaneous $e$-ph
interactions. Our treatment resolves these issues by providing
a nonperturbative solution to the original $e$-ph
model at a physically motivated phonon frequency 
allowing for retardation. What remains now is to push these methods into the low-temperature metallic phase of the doped system. 

In summary, we have focused on the parent compound of the bismuthate families of superconductors and obtained nonperturbative results for a model derived from {\it ab initio} electronic structure calculations. 
This is in contrast to more traditional QMC studies of these systems, which have traditionally focused on model Hamiltonians with large phonon energies $\hbar\Omega_0/t \approx 0.5-1$~\cite{ScalettarPRL1989, WeberPRB2015,  EsterlisPRB2018, LiQM2020, NosarzewskiPRB2021, CaiPRL2021, CostaPRL2021, XingPRL2021}. Our work, therefore, represents a crucial bridge between  {\it ab initio} nonpertubative numerical simulations of quantum materials with strong $e$-ph interactions. In this spirit, our results mark an important initial step to understanding a much broader range of perovskite materials involving the bond-stretching motion of their ligand anions. These include other negative charge-transfer materials like the rare-earth nickelates  $R$NiO$_3$~\cite{ParkPRL2012,JohnstonPRL2014} or the various families of $sp$-$ABX_3$ perovskites~\cite{BenamPRB2021}.

\section{Methods}

\subsection{The non-interacting band structure}

In the absence of $e$-ph coupling, the Hamiltonian can be diagonalized exactly in momentum space. Introducing Fourier transforms $s^\dagger_{{\bf i},\sigma} = \frac{1}{\sqrt{N}}\sum_{\bf k} e^{-\mathrm{i}{\bf k}\cdot{\bf R}_i}s^\dagger_{{\bf k},\sigma}$ and $p^\dagger_{{\bf i},\delta,\sigma} = \frac{1}{\sqrt{N}}\sum_{\bf k} e^{-\mathrm{i}{\bf k}\cdot({\bf R}_i+\hat{\delta}a/2)}p^\dagger_{{\bf k},\delta,\sigma}$ for the Bi $6s$ and O $2p_\gamma$ orbitals, we have $H_0 = \sum_{{\bf k},\sigma}\psi^\dagger_{{\bf k},\sigma}H^{\phantom\dagger}_{\bf k}\psi^{\phantom\dagger}_{{\bf k},\sigma}$, where $\psi^\dagger_{{\bf k},\sigma} = [s^\dagger_{{\bf k},\sigma}, 
p^\dagger_{{\bf k},x,\sigma}, p^\dagger_{{\bf k},y,\sigma}, p^\dagger_{{\bf k},z,\sigma}]$ and 
\begin{equation}
    H_{\bf k} = \left[\begin{array}{cccc}
    \epsilon_s -\mu & \mathrm{i}t_x({\bf k}) & \mathrm{i}t_y({\bf k})&  \mathrm{i}t_z({\bf k})  \\
    -\mathrm{i}t_x({\bf k}) & \epsilon_p-\mu & t^\prime_{x,y}({\bf k}) & t^\prime_{x,z}({\bf k}) \\
    -\mathrm{i}t_y({\bf k}) & t^\prime_{x,y}({\bf k}) & \epsilon_p-\mu & t^\prime_{y,z}({\bf k}) \\    
    -\mathrm{i}t_z({\bf k}) & t^\prime_{x,z}({\bf k}) & t^\prime_{y,z}({\bf k}) & \epsilon_p-\mu 
    \end{array}\right], 
\end{equation}
with $t_\gamma({\bf k}) = -2t_{sp}\sin(k_\gamma a/2)$ and $t^\prime_{\delta,\delta^\prime}({\bf k}) = 4t_{pp}\sin(k_\delta a/2)\sin(k_{\delta^\prime}a/2)$. \\

\subsection{The dimensionless electron-phonon coupling}

The dimensionless electron-phonon coupling constant for an arbitrary phonon branch $\Omega({\bf q})$ that is relevant in theories of superconductivity is given by 
$\lambda = \frac{1}{N}\sum_{\bf q} \lambda_{\bf q}$, where $\lambda_{\bf q}$ is a mode-resolved coupling defined as 
\begin{align}\label{eq:lambda2}
    \lambda_{\bf q} = \frac{2}{N(0)N} \sum_{\bf k} \frac{|g({\bf k},{\bf q})|^2}{\hbar\Omega(\bf q)}\delta(\epsilon_{\bf k}-\mu)\delta(\epsilon_{{\bf k}+{\bf q}}-\mu).
\end{align}
Here, $N(0) = \frac{1}{N} \sum_{\mathbf{k}} \delta(\epsilon_{\mathbf{k}})$ is density of states at the Fermi level per spin species, and $g({\bf k},{\bf q})$ is the $e$-ph vertex for scattering within the band crossing the Fermi level 
$g(\mathbf{k},\mathbf{q}) = \boldsymbol{\phi}_{0,\mathbf{k}+\mathbf{q}}^\dagger
    G({\bf k},{\bf q})
    \boldsymbol{\phi}^{\phantom\dagger}_{0,\mathbf{k}}$,
where 
\begin{equation}
    G({\bf k},{\bf q})=\left[\begin{array}{cccc}
        0 & g_{x}^{*}({\bf k}+{\bf q}) & g_{y}^{*}({\bf k}+{\bf q}) & g_{z}^{*}({\bf k}+{\bf q})\\
        g_{x}(\mathbf{k}) & 0 & 0 & 0\\
        g_{y}(\mathbf{k}) & 0 & 0 & 0\\
        g_{z}(\mathbf{k}) & 0 & 0 & 0
    \end{array}\right].  
\end{equation}
Here $g_{\delta}(\mathbf{k}) = 2g\cos\big(\tfrac{k_{\delta} a}{2}\big)$, and 
$g=\sqrt{(\hbar\alpha_{sp}^2)/(2M\Omega_0)}=99.4$~meV. In the above, $\boldsymbol{\phi}_{0,\mathbf{k}'}$ is the vector appearing in the Bloch wave function $\boldsymbol{\psi}_{0,\mathbf{k}'}(\mathbf{r})=\boldsymbol{\phi}_{0,\mathbf{k}'} e^{{\rm i}\mathbf{k}'\cdot\mathbf{r}}$ for the band crossing the Fermi level, and is solved for by numerically diagonalizing $H_\mathbf{k}$.

To calculate $\lambda$ in the noninteracting limit, we set $\Omega({\bf q}) = \Omega_0$ and approximate the delta functions appearing in Eq.~\eqref{eq:lambda2} with  Lorentz distributions $\delta(x) \approx L_\gamma(x)$ and half-width at half-maximum set to $\gamma = 10$~meV.We then calculate the average over the FS using a $50^3$ grid of momentum points. (We have checked that our results are well converged for these values.) This procedure results in $\lambda = 0.18$.
The corresponding value of the mode-resolved coupling $\lambda_{\bf q_\mathrm{cdw}} = 0.335$. Interestingly, these values are comparable to early \gls*{DFT} estimates for coupling to the bond-stretching modes \cite{MeregalliPRB1998}, as well as the value for the total coupling extracted from  experiments on the bismuthates~\cite{GraebnerPRB1989, MarsiglioPRB1996}. If we instead neglect the momentum dependence in the vertex and apply the approximations $g(\mathbf{k},\mathbf{q}) \approx g$ and $N(0) \approx 1/W$, where $W=5.62$~eV is the bandwidth of the lowest energy band, the result is much smaller, with $\lambda \approx 0.0586$.

To estimate the effective total $\lambda^\text{eff}$ and mode-resolved $\lambda_{\bf q}^\text{eff}$ dimensionless coupling constants, we approximated $\Omega_{\bf q}$ in Eq.~\eqref{eq:lambda2} with the renormalized dispersion $\Omega({\bf q},0)$ [Eq.~\ref{eq:omega_dressed}]. In this case, we interpolated the results for the $L = 10$ lattice onto the $50^3$ grid of momentum points using a sinc interpolation scheme \cite{FourierTools.jl}. Using this approach, we obtain $\lambda^\text{eff} = 0.22$ and $\lambda^\text{eff}_{\mathbf{q}_{\rm cdw}} \approx 9.92$, respectively, at $\beta = 17$ eV$^{-1}$. The enhancement in the mode-resolved coupling at $\mathbf{q}_{\rm cdw}$ can be attributed to the phonon softening, with $\lambda^\text{eff}_{\mathbf{q}_{\rm cdw}}/\lambda_{\mathbf{q}_{\rm cdw}} \approx [\Omega(\mathbf{q}_{\rm cdw},0)/\Omega_0]^{-1}.$\\

\subsection{An RPA estimate for $T_\mathrm{cdw}$}

The transition temperature to the \gls*{CDW} is signaled by a divergence in the charge susceptibility $\chi({\bf q})$ in the thermodynamic limit. For weak coupling, $\chi({\bf q})$ can be approximated using the RPA and calculated from the infinite sum polarization bubble diagrams to yield \cite{EiterPNAS2013}
\begin{equation}\label{eq:RPA_chi}
    \chi_\mathrm{RPA}({\bf q}) = 
    \frac{\chi_0({\bf q})}{1+D_0({\bf q},0) \chi_{g,0}({\bf q})}. 
\end{equation}
Here, $\chi_0({\bf q})$ is the bare (Lindhard) susceptibility defined in the main text, $D_0({\bf q},0) = -2/\hbar\Omega_0$ is the noninteracting phonon Green's function at zero frequency, and $\chi_{g,0}({\bf q})$ is shorthand for   
\begin{eqnarray}
    \chi_{g,0}({\bf q}) = -\frac{2}{N}
    \sum_{{\bf k}} |g({\bf k},{\bf q})|^2\frac{f(\epsilon_{{\bf k}+{\bf q}})-f(\epsilon_{\bf k})}{\epsilon_{{\bf k}+{\bf q}}-\epsilon_{\bf k}},
\end{eqnarray}
with $f(x) = [1+\exp(\beta x)]^{-1}$ the usual Fermi factor. The \gls*{CDW} transition thus occurs when the denominator of Eq.~\eqref{eq:RPA_chi} goes to zero. Using a set of $50^3$ momentum points, $\chi_{\rm RPA}(\mathbf{q}_{\rm cdw})$ diverges at $T_{\rm cdw} = 2630$~K, over a factor of two larger than the value obtained from HMC.\\

\subsection{HMC simulation details}

As previously discussed, we used a recently introduced collection of algorithms to perform the numerical simulations described in this work. Below we highlight some of the relevant parameter values used in the simulation and refer the reader to Ref.~\cite{CohenSteadPRE2022} for more information on their definitions.

The \gls*{HMC} approach employed in this work is based on a path-integral formulation of the quantum partition function. For this, we used an imaginary-time discretization of  $\Delta\tau = 0.05 \ {\rm eV}^{-1}$. With this choice, global discretization errors of order $\mathcal{O}(\Delta \tau^2)$ are well controlled. Updates to the phonon field were proposed using \gls*{HMC} trajectories of $N_t=10$ time-steps, and accepted or rejected according to a Metropolis criterion that ensures detailed balance. We solve for the force in our effective dynamics using the conjugate gradient method with a relative residual error threshold of $\epsilon_{\rm max}=10^{-5}$. For the Metropolis accept/reject step, we reduce the threshold to $\epsilon_{\rm max}=10^{-10}$.

Forces in the HMC dynamics can be decomposed into fermionic and bosonic parts. The former are more expensive to calculate, but also of smaller magnitude, so we use them to take relatively large time-steps of $\Delta t = 2$. Within each of these large time-steps, we employed a time-step splitting algorithm that involves $n_t=100$ sub time-steps. The much smaller time-step value $\Delta t' = 0.02$ accounts for the rapidly evolving bosonic forces. To reduce autocorrelation times further, we apply Fourier acceleration via a dynamical mass matrix (regularization parameter $m_{\rm reg}=0.5$) that modifies the effective relaxation time scale according to the imaginary-time wavelength. Systems were initially thermalized using $N_{\rm therm}=1000$ HMC trial updates. After thermalization, an additional $N_{\rm sim}=4000$ updates were performed. During this simulation stage, measurements were taken after every HMC trial update. Each measurement employed $N_{\rm rv}=10$ random vectors to estimate the Green's function and higher-order correlation functions.

\section*{Data Availability}

The data that support the findings of this study are accessible as a public repository at \url{https://doi.org/10.5281/zenodo.7516925}. 

\section*{Code Availability}
The primary Julia based simulation code used to generate the results presented in this paper is accessible as a public repository, \url{https://github.com/cohensbw/ElPhDynamics.git}. Instruction on how to use the code will be provided upon reasonable request.

\section*{Acknowledgements} 
The authors thank M. Berciu, S. Li, and G. A. Sawatzky for useful discussions. This work was supported by the U.S. Department of Energy, Office of Science, Office of Basic Energy Sciences, under Award Number DE-SC0022311. 

\section*{Author Contributions}

S.J. conceived of the project and supervised the research. B.C-S., K.B. and R.S. developed the code. B.C-S. collected the results and performed the calculations. B.C-S., S.J., K.B. and R.S. discussed the results. B.C-S. and S.J. wrote the manuscript.

\section*{Competing Interests}

The authors declare no competing  financial or non-financial interests.

\end{document}